\documentclass[manuscript]{aastex}
\usepackage{txfonts}
\usepackage{graphicx}

\slugcomment{APRIM proceeding}

\shorttitle{Inflows in massive star formation regions} \shortauthors{Wu et al.}

\begin{document}

\title{Inflows in massive star formation regions}
\author{Yuefang Wu            \altaffilmark{1}
,Tie Liu               \altaffilmark{2},
,Shengli Qin          \altaffilmark{3  }
}

\altaffiltext{1}{Peking University, China; yfwu.pku@gmail.com}
\altaffiltext{2}{Korea Astronomy and Space Science Institute, Korea; liutiepku@gmail.com}
\altaffiltext{3}{YunNan University, China}

\begin{abstract}
How high-mass stars form remains unclear currently. Calculation
suggests that the radiation pressure of a forming star can halt
spherical infall, preventing its further growth when it reaches 10
M$_{\odot}$. Two major theoretical models on the further growth of
stellar mass were proposed. One model suggests the mergence of
less massive stellar objects, and the other is still through
accretion but with the help of disk. Inflow motions are the key
evidence of how forming stars further gain mass to build up
massive stars. Recent development in technology has boosted the
search of inflow motion. A number of high-mass collapse candidates
were obtained with single dish observations, mostly showed blue
profile.
The infalling signatures seem to be more common in regions with
developed radiation pressure than in younger cores, which opposes
the theoretical prediction and is also very different from that of
low mass star formation. Interferometer studies so far confirm
such tendency with more obvious blue profile or inverse P Cygni
profile. Results seem to favor the accretion model. However, the
evolution tendency of the infall motion in massive star forming
cores needs to be further explored. Direct evidence for monolithic
or  competitive collapse processes is still lack. ALMA will enable
us to probe more detail of gravity process.
\end{abstract}

\keywords{stars: formation---stars: pre-main sequence---ISM: kinematics and
dynamics}

\section{Introduction}

Gravitational collapse is an essential process for star formation.
However, when the forming stellar object reaches 10 M$\odot$, the
strong radiation pressure can halt material falling (Wolfire \&
Cassinelli 1987). A new model that more massive stars can be
formed through coalescence of less massive stellar objects was
suggested (Bonnell et al. 1998). The accretion model was also
improved (Yorke \& Sonnhalter 2002; Jijina \& Adams 1996).
Observational evidences of material inward motions in massive star
formation regions are critical to test these models. Nevertheless
it is more difficult to seek material infalling in massive star
formation regions than in low mass ones because of their complex
environment, large distance and quick evolution. The interactions
between environment of massive young stellar objects and their
feedback bring additional difficulties for the identification of
the gas infall motion. The development of millimeter and
sub-millimeter equipments made such probing as feasible. In the
recent decade not only a number of searches for the gravitational
evidence in high mass star formation regions were carried out by
single dishes, but typical sources were also examined with
interferometers. These have greatly deepened our understanding of
massive star formation. Since some basic questions still remain,
more observations are needed in the future.

\section{Surveys of inflow motions in massive star formation regions}

Ten years after the blue profile detected in the low mass core of
B335 (Zhou et al. 1993) it was found in a spectroscopic survey
toward 28 massive star formation cores associated with H$_2$O
masers (Wu \& Evans 2003). With CSO and IRAM,  HCN (3-2), CS
(5-4), (3-2), (2-1), and  H$^{13}$CN (3-2) were used as optically
thick and thin lines, blue profiles were obtained in 12 cores and
red profiles in 6 cores. Soon 77 candidates of high mass
protostellar objects (HMPOs) were searched with IRAM and JCMT
(Fuller et al. 2005). They identified 22 promising infall
candidates. Meanwhile toward 12 UC~H{\sc ii} regions, Wyrowski et
al (2006) detected 9 blue profiles with CO (4-3) and $^{13}$CO
(8-7) lines by APEX.  Using MOPRA, Purcell et al. (2006) observed
83 CH$_3$OH maser-selected regions with lines of CH$_3$OH (5-4),
(6-5), and HCO$^+$ and H$^{13}$CO$^+$ (1-0) transitions. They
detected 12 blue profiles. Klaassen \& Wilson (2007) observed 23
UC~H{\sc ii} regions with outflows using JCMT. They used tracers
of HCO$^+$ (4-3), H$^{13}$CO$^+$ (4-3), C$^{}$O (2-1) and found 9
sources having Infall motion. Toward very early Orion cores which
are likely precursors of protostars, Velusamy et al.(2008)
detected 27 Orion cores with HCO$^+$ and H$^{13}$CO$^+$ (3-2) by
CSO. They found dichotomy in the dynamical status: 9 sources have
blue profile and 10 have red profile. These surveys obtained a
number of inflow candidates and show inflow motions are common in
massive star formation regions.

To further examine characteristics of inflow motions in massive
star formation regions, a mapping survey of  HCO$^+$(1-0),
CS(3-2), N$_2$H$^+$(1-0), C$^{18}$O(1-0) was made with IRAM (Wu et
al. 2007). Rotation could be excluded from the spatial
distribution of the asymmetric line profiles. Also the peak
positions of blue profile and associated molecular outflow can be
identified. To see difference of inflow motion at different
evolution status, this survey includes two group sources. Group I
contains 33 UC~H{\sc ii} precursors (PUC~H{\sc ii} or HMPOS) and
Group II consists of 12 UC~H{\sc ii} regions. Using HCO$^+$(1-0)
nine and seven blue profiles were obtained in Group I and Group II
respectively. However there are 4 red profiles in Group I while no
red profile in Group II. The asymmetric lines detected result in
the blue excess E= is 0.17 and 0.58 for the two groups
respectively, here
 E= (N$_{B}$-N$_{R}$/N$_T$), N$_{B}$, N$_{R}$ and N$_T$ are the number
of sources with blue profile, red profile and the sample of the
survey (Mardones et al. 1997). Results show the UC~H{\sc ii}
regions have higher blue excess than their precursors. Meanwhile
the results of the survey shows blue profiles are usually peaked
at the core center and part of the sources have high velocity
outflows. Figure 1 presents the HCO$^+$ (1-0) mapping grid,
comparing of optical thin and thick lines as well as P-V diagrams
of two sources, one is a PUC~H{\sc ii} and the other is associated
with an UC~H{\sc ii} region.

We compare the blue excess of the two group sources of this survey
with those of previous surveys. Table 1 gives the blue excess (E)
of cores at different evolution statues. We can see the following
situations:

(1) For the same  HMPOS, With the same tracer HCO$^+$ (1-0), the E
of  two surveys are about the same, $\sim$ 16\%;

(2)  Using different tracers of HCO$^+$(1-0) and CO(4-3), E$>$50\%
was obtained for UC~H{\sc ii} regions;

(3) For the youngest massive cores (see Table 1), the E value is
negative.

We also compare the E of massive star formation regions with their
low mass counterparts. From Table 1, one can see that  there is no
tendency of E varying with time for the low mass cores.  However
the tendency that E of the late phase is larger than that of the
early phase for massive star formation cores is evident. This
contradicts with the theoretical result since radiation pressure
of UC~H{\sc ii} regions should be larger than that of earlier
phases. Possible explanations for the difference may be concerned
with thermalization of the flow regions, influence of outflow or
turbulence and gas reserves (Wu et al. 2007), which need to be
further explored.

\section{Probing inflows with high angular resolution observations}

To examine inflow motions in the inner regions of massive cores
and test the results of the single dish observations, higher
angular resolution observations are needed. Especially
interferometer is powerful to probe deep layers of cores. Recent
years excellent examples such as W51N, NGC 7538 and Sgr B2 were
obtained (Zapata et al. 2008; Qin et al. 2008; Qiu et al. 2011).
Some other massive core collapse candidates from single dish
observations were observed with various interferometers. Below we
make a briefly comparison of the line signatures of these cores
observed with single dishes and interferometers. Table 2 gives the
source name, evolution phase, signatures detected with different
resolution observations. The comparing results are as following:

(1) Generally no obvious confliction was found between the results
of single dish and interferometer observations in these cores. And
more strong signature of gravitational collapse were detected with
interferometers. Core JCMT18354-0649S was mapped with  HCN,
HCO$^+$, H$^{13}$CO$^+$ (3-2) and  C$^{17}$O (2-1) at JCMT  (Wu et
al. 2005). Blue profile was shown in multiple pairs of optical
thick and thin lines. Two layer model fitting gave infall velocity
of 0.25 km/s. The SMA observations of molecular lines including
CH$_3$OH (5$_{23}$-4$_{13}$) and HCN (3-2) also showed blue
profile which is more prominent than the one obtained by JCMT (Liu
et al. 2011a). The infall velocity is 1.3 kms$^{-1}$ from the same
model fitting. The inflow signature of W3-SE obtained with IRAM
was confirmed by the Combined Array for Research in
Millimeter-wave Astronomy (CARMA) (Zhu et al. 2010 ). For NGC7538,
blue profile was observed with single dish, while inverse profile
 was obtained with SMA (Zhu et al. 2013: Qiu et al. 2011).

(2) Profiles were distinguished in inner regions of the cores. For
example, high excitation density molecular lines
 of CH$_3$CN(12$_4$-11$_4$) and CH$_3$OH(8$_{-1,8}$-7$_{0,7}$)
detected with the Atacama Large Millimeter/Submillimeter Array
(ALMA) in Orion KL region present collapse signature for the first
time, which show an inverse P Cygni profile for Source I and blue
profile for the hot core (Wu, Liu \& Qin 2014). In G9.62-0.19,
there are compact cores C, D, E, F and I within a region with a
diameter $<$ 5$^{\prime\prime}$  (Testi et al. 2000). Core E is a
young massive star surrounded by a small UC~H{\sc ii} region
(Hofner 1996), and core F is a very young stellar object (Linz et
al. 2005). Hofner et al. (2001) observed G9.62-0.19 with HCO$^+$
(1-0), SO (43-32) and SiO (5-4) lines using IRAM. The HCO$^+$
(1-0) line shows both blue and red shifted absorption at a spatial
resolution of 27$^{\prime\prime}$. SMA observation revealed that
the blue asymmetric profile comes from Core E and the red one from
Core F (Liu et al. 2011b ).  The profile difference of the two
cores Source I and the hot core in Orion KL, and core E and core F
in G9.62-0.19 is consistent with the results of single dish
surveys, E$_{Late}$ $>$ E$_{Early}$.

(3) Infall velocity difference at different size of inflow regions
was found. For example, in G19.62-0.23, the red shifted gas
absorption region of the CN (N=3-2) line is smaller than that of
$^{13}$CO (3-2), while the infall velocity obtained from CN (3-2)
line is larger than that from $^{13}$CO (3-2), which is consistent
with inside out collapse model (Shu, Adams \& Lizano. 1987). G34.
26+0.15 is the only source that inverse P Cygni profile was
detected in the IRAM survey (Wu et al. 2007). SMA detected six
cores in this region. Multiple CN (N=2-1) lines of each core show
inverse P Cygni profile (Liu et al. 2013a). The largest infall
velocity corresponds to the smallest absorption area. The tendency
of the velocity changing with the inflow region size is consist
with the inside-out model too.

(4) Fragmentation was detected. In G10.6-0.4, there are seven
sub-millimeter cores. All cores show red shifted absorption gas
(Liu et al. 2013b), four of which are associated with infrared
point sources. The center core has largest mass and largest infall
velocity. The difference of the kinematic and Bondi-Hoyle mass
accretion rates is agreed within a factor of 2. These results are
consistent with competitive accretion (Bonnel et al. 2001).
However, the core structure and mass spectrum need to be further
examined.

(5) The SMA observations of  G8.68-0.37 did not show  infall
motion signatures. However, the lines of CO (1-0) and (2-1)
 observed with PMO and CSO show prominent blue profile. While the
HCN(3-2) lines observed with JCMT present red profile although the
S/R ratio is low. This source contains an EGO (Ren et al. 2012).
Similar situation may be happened in some other EGOs. The survey
toward 72 EGO by PMO (Chen at al. 2010) found 29 sources having
blue profiles, and 19 have red profiles, giving E of 0.14. In
another HCN (3-2) survey of EGOs (Wu, Liu, Reipurth, on going)
with JCMT, the blue asymmetric and red asymmetric profiles are 8
and 14 respectively, giving -0.24 of E. In cores like G8.68-0.37,
there are inflow motions outside and expanding in the inner, which
 needs to be studied by more observations. In source G45.12+0.13, NH$_3$ (4,4) and
(5,5) lines detected with 100 m telescope at Effelsberg show
inverse P Cygni profiles (Cesaroni, Walmsley \& Churchwell 1992).
The VLA observations of NH$_3$ (2,2) and (4,4) lines with beam
size 2$^{\prime\prime}$.9$\times$2$^{\prime\prime}$.7  also showed
such profiles (Hofner, Peterson,\& Cesaroni 1999). However, high
spatial resolution observations of 5$^{\prime\prime}$ failed to
identify inverse P Cygni profile in this region \citep{wil96}.
Such intricate situation in the inner region of a source needs to
be explored too.

\begin{table*}[t!]
\caption{Blue excesses of inflow surveys\label{tab:pkastable1}}
\centering
\begin{tabular}{ccccc}
\cline{1-5}
\cline{1-5}
Sources        &  & Evolutionary phases &  & Ref.  \\
High mass       &  Earlier than PUC~H{\sc ii} &       PUC~H{\sc ii}        &      UC~H{\sc ii}                  \\
  examples      &                      &     Core JCMT       &      G 34.26   & 1,2 \\
 E (HCO+(1-0))  &             …        &   17\%              &  58\%           & 3 \\
                &                      &   15\%              &  70\% (CO 4-3)             &4,5  \\
    HCO+(3-2)   &     -0.04            &                     &                             &,6 \\
\cline{1-5}
Low mass        &   Class -I           &       Class 0      &             Class I          \\
   examples     &   L1544              &     B335            &         L1251B          &7,8     \\
 E HCN (3-2)    &   30\%               &       31\%          &            31\%         &8     \\
\cline{1-5}
\cline{1-5}
\end{tabular}
\\
\noindent Ref.: 1, Liu et al. 2011a; 2. Liu et al. 2013;
3. Wu et al. 2007; 4. Fuller et al. 2005; 5. Wyrowski et al. 2006;
6. Velusamy et al. 2008; 7. Mardones et al. 1997 ; 8. Evans (2003)
\end{table*}


\begin{table*}[t!]
\caption{A comparison between single dish and interferometer
observed results\label{tab:pkastable1}} \centering
\begin{tabular}{ccccc}
\cline{1-5}
\cline{1-5}
Source      &    Phase      &   Single dish  &     High resolution  &  Ref.      \\
\cline{1-5}
JCMT 18354       &  PUC~H{\sc ii}             &   Blue             &Blue             & 1.2         \\
   W3-SE         &  PUC~H{\sc ii}             &   Blue             &Blue             & 3,4         \\
G9.62-0.19F      & Younger PUC~H{\sc ii}      &    Blue or Red?    &    Red          &    5.6      \\
Orion KL/Hot     & core Younger PUC~H{\sc ii} &    ---             & Blue            &   7         \\
   G8.68         &   PUC~H{\sc ii}            & Blue-outer; Red-iner? &    --        &     8       \\
 G19.61-0.23     &     UC~H{\sc ii}          &     Blue or Red?   & Inverse P Cygni &     3,9     \\
G9.62+0.19E      &   UC~H{\sc ii}            &   Blue or Red?     &   Blue          & 5,6         \\
 NGC7438 IRS1    &   UC~H{\sc ii}            &       Blue         &Inverse P CYgni  &   3, 10     \\
    G10.6-0.4    &   UC~H{\sc ii}          &         Blue       &Inverse P Cygni  &   11,12     \\
  G34.26-0.15    &   UC~H{\sc ii}          &    Inverse P Cygni &  Inverse P Cygni&     3, 13   \\
 Orion KL/Source I &  Radio source     &   ---              & Inverse P Cygni &      7      \\
 G45.12+0.13     &     UC~H{\sc ii}          &    Inverse P Cygni & Inverse P Cygni &    14, 15   \\
                 &                     &                    & No inverse PCygni&    16      \\
\cline{1-5}
\cline{1-5}
\end{tabular}
\\
\noindent Ref.: 1. Wu et al. 2005; 2. Liu et al. 2011; 3.
Wu et al. 2007; 4. Zhu et al. 2010; 5. Hofner et al. 2001; 6. Liu
et al. 2011; 7. Wu et al. 2014;  8. Ren et al. 2012; 9. Wu et al.
2009; 10. Zhu et al. 2013; 11. Wu \& Evans, 2003; 12. Liu et al.
2013a; 13. Liu et al. 2013b
\end{table*}

\section{Summary and prospect}

Searches for evidence of gravitational collapse of massive star
formation cores have been progressed greatly. During the last
decade after blue profile found in low mass cores, searches toward
various phases of massive dense cores were frequently made.
Results suggest that inflow motions are common in massive star
formation regions too.

A number of the candidates were observed with high angular
resolution and strong evidence of inverse P Cygni was found.
Velocity changes at different sizes of inflow regions were
detected. Multiple cores and their inflow motions were also
obtained. These results suggest that  high mass stars may form
via accretion model.

There may be foundational difference between collapse of the two
kinds of star formation processes. Surveys in massive cores at
different evolutional statuses show that there is evolution
tendency of the blue excess. And profiles of core pairs which is
consisted of a young massive core and a UC~H{\sc ii} region are
consistent with E$_{Late}$ $>$ E$_{Early}$. Inflow motion appears
to occur at outer side while expanding or outflow occur inside of
some massive cores, which is not seen in low mass cores so far.
Structure, mass distribution and inflow processes of massive dense
cores need more probing. ALMA has better sensitivity and
resolution, which will enable us detect fragmentation or mass
secretion, and determine their mass spectra. Future studies will
provides us better pictures of gravitational collapse in massive
star formation regions.

\begin{figure}[t]
\centering
\includegraphics[angle=-90,width=90mm]{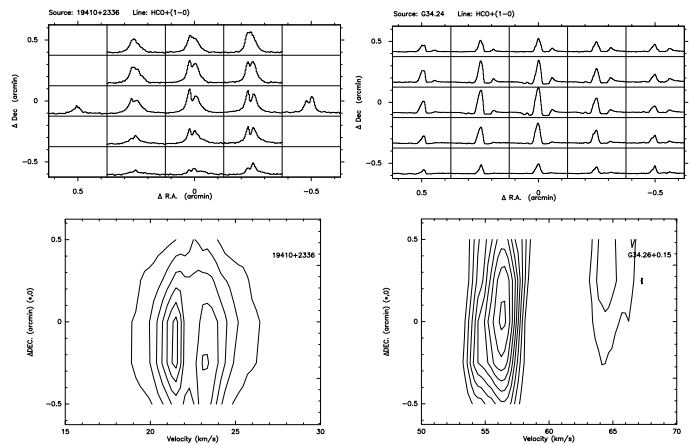}
\caption{Left: UC~H{\sc ii} precursor 19410+2336; Right: UC~H{\sc
ii} region G34.26+0.15. (See Table 1 of Wu et al. 2007. The
figures are not published yet.) \label{fig:pkasfig1}}
\vspace{15mm} 
\end{figure}


\acknowledgments This work was supported by the China Ministry of
Science and Technology under State Key Development Program for
Basic Research (2012CB821800), the grants of NSFC number
11373009,11373026, 11433004, 11433008 and Midwest universities
comprehensive strength promotion project (XT412001, Yunnan
University).


{}


\end{document}